\documentclass[a4paper,11pt]{article}
\pdfoutput=1 

\usepackage{jheppub} 

\usepackage[normalem]{ulem}	
\usepackage[latin1]{inputenc}
\usepackage{amsfonts}

\title{\boldmath
Neutrino hierarchy and fermion spectrum from a single family in
six dimensions: realistic predictions }

\author[a]{J.-M.~Fr\`ere,}
\author[b,c]{M.~Libanov,}
\author[a]{S.~Mollet}
\author[b]{and S.~Troitsky}

\affiliation[a]{
Service de Physique Th\'eorique, Universit\'e Libre de Bruxelles,\\
Campus de la Plaine CP225, Bd du Triomphe, 1050 Brussels, Belgium
}
\affiliation[b]{
Institute for Nuclear Research of the Russian Academy of Sciences,\\
60th October Anniversary Prospect 7a, 117312, Moscow, Russia
}
\affiliation[c]{
Moscow Institute of Physics and Technology,\\
Institutskii per., 9, 141700, Dolgoprudny, Moscow Region, Russia
}

\emailAdd{frere@ulb.ac.be}
\emailAdd{ml@ms2.inr.ac.ru}
\emailAdd{smollet@ulb.ac.be}
\emailAdd{st@ms2.inr.ac.ru}

\abstract{%
In this paper, we move from a ``proof of concept'' to challenging
predictions in a flavour model emerging from a single fermion family in six
dimensions (6-D) with the two extra dimensions compactified on a
sphere. The most striking predictions (beyond the basic ingredients
already demonstrated for a realistic quark mass pattern) are for  the
neutrino sector. Not only can 4-D Majorana masses be generated from 6-D
(where no Majorana spinors exist), but Majorana masses are naturally
associated to large mixing. The model favours inverted hierarchy and a
quasi-Dirac partial suppression of neutrinoless double beta decay.}

\keywords{Large Extra Dimensions, Quark
Masses and SM Parameters, Neutrino Physics}
\preprint{\parbox[t]{3cm}{INR-TH-2013-27%
}}

\begin{document}
\maketitle
\flushbottom

\section{Introduction}
\label{sec:intro}

In the recent couple of years, the particle physics was faced with two
important experimental discoveries. One is the new boson with mass about
125~GeV \cite{boson:ATLAS, boson:CMS} whose properties look now very
similar \cite{boson:ATLAS:properties, boson:CMS:properties} to those of
the Standard Model Scalar boson required by the Brout-Englert-Higgs
mechanism \cite{BroutEnglert, Higgs}. The other one is the nonzero mixing
between the first and the third neutrino generations whose value was
determined in several experiments, with growing precision \cite{K2K,
DayaBay}. While the first one represents a triumphal support to the
Standard Model (SM), the second one reminds us how little we understand
the flavour sector. Neutrino mixing is impossible in the SM where neutrinos
are strictly massless. While neutrino masses can in principle be added to
the SM with the help of right-handed neutrinos and very small Yukawa
couplings, the alternate possibility (Majorana or Weyl neutrinos) is quite
attractive. However, we still do not have a clue to the physical
realization of this mechanism.

One of the approaches, which allows to explain naturally the hierarchy of
masses and mixings both in the quark and lepton sectors but does not
necessarily require new physics at the LHC scale, explores the idea of
extra space dimensions with the SM matter localized on a brane or a
topological defect. We have shown in previous papers how the topological
structure can be chosen in a 6-D model for existence of three effective
four-dimensional modes for each multidimensional fermion. (Three is for
the moment an arbitrary -- empirically motivated --  choice; we will
however argue that four families would be difficult to fit within the
experimental constraints). Thus a single generation of fermions exists in
the fundamental theory, with a corresponding reduction of the number of
free parameters. Meanwhile, the correct pattern of masses and mixings is
generated without fine tuning by overlaps of the wave functions. This
rigidity of the framework is strikingly different from
``multilocalisation'' schemes, where various fermions are located on
branes with tuned position in the 5th dimension. A working example with
six space-time dimensions and the Abelian vortex as a brane has been
constructed \cite{LT, FLT} and elaborated \cite{FLT_neutrino, LN_fit,
LN_fit_t, FLNT_sphere, FLNT_flavour, FLNT_LHC, LN_Higgs, LN_Higgs_Yad}. We
have later shown that the very same 6-D mechanism, can provide 4-D
Majorana (Weyl)  neutrinos, an astonishing fact in itself, since Majorana
fermions don't exist in 6-D. This surprising possibility comes with
tantalizing implications: it implies naturally large mixings and an
inverted hierarchy for masses in the neutrino sector \cite{FLL_neutrino}
(see refs.~\cite{Libanov:2011zz, Libanov:2011st}).

In this note, we go beyond the ``proof of concept'', further discuss  the
latest developments in our six-dimensional approach to the flavour
structure of the SM, but also provide a convincing fit to the mass and
mixing spectra. Beyond this specific fit, we also outline what we consider
to be the ``generic'' features of the model, and suggest possible tests.
We also consider briefly (in section~\ref{sec:experiment})  the
possibility to study these key features in an ad-hoc effective 4-D
approach, in particular for test of flavour-changing but family-number
conserving interactions. This allows to put purely experiment-based limits
for possible detection of flavour-carrying new particles at LHC.

While we had already stressed some of these points (notably the need for a
light Standard Model Scalar~\cite{LN_Higgs, LN_Higgs_Yad}), the latest
data allow to
narrow the fit. In terms of quarks, we reproduce the mixing structure with
a minimal number of parameters, and obtain decent values for the masses.
In terms of neutrinos, large mixing (for Majorana neutrinos in four
dimensions) appear naturally~\cite{FLL_neutrino}. While we  already
announced ({\it in tempore non suspecto})  a finite value of $\theta_{13}$
(now seen to be compatible with experiment), we confirm the distinctive
prediction of inverted hierarchy and nearly pseudo-Dirac mass matrix,
which minimizes the neutrinoless double beta decay (that is, its value is
found at the very bottom of the inverted hierarchy allowed band). CP
violation in neutrino oscillations is also expected.

One major characteristic is the presence of some family-number symmetry.
In the absence of CKM this is indeed exact, and leads to conservation of
``family number''. We have shown previously that this ``horizontal
symmetry'', mediated by Kaluza-Klein (KK) excitations, is perfectly
compatible with existing data, but puts a lower bound on the 6-D scale. It
provides however for interesting phenomenology, and may be a way of
testing the model. Note that these ``horizontal'' interactions are not
directly responsible for the mass generation, and thus avoid the major
conflict with experiment seen, for instance, in many ``extended
technicolor'' models. The $\sim50$ TeV lower bound (obtained from limits
on flavour-changing processes like $K_L^0 \rightarrow \bar{e} \mu $)
results from the estimations of wave functions overlaps in the specific
case considered (compactification on a sphere). It is also possible to
write the interaction in a way independent of these overlaps, in an
effective approach. In that case, possible discovery at LHC can be related
to the precision tests in a model-independent way.

The paper is organized as follows. We briefly review the construction of
the model in section~\ref{sec:model}. Then, we demonstrate explicitly how
the observed value of the Standard Model Scalar mass fits the overall
construction and fix the value of parameters in the scalar sector
(section~\ref{sec:scalar}). A possible Goldstone mode is avoided through
the Yukawa couplings. We present the fermionic interactions and fit masses
and mixing of the matter fields in section~\ref{sec:fermion}. This
successful simultaneous fit of parameters related to scalar,
charged-fermion and neutrino sectors allows us to proceed with an update
on quantitative experimental predictions, distinctive for our model (and a
possible model-independent approach), in section~\ref{sec:experiment},
while we conclude briefly in section~\ref{sec:concl}, and mention some
possible alternatives to the choices made in this paper.

\section{The model: a general description}
\label{sec:model}

The basic model has been presented in successive forms, first \cite{LT}
with a minimal number of fields in flat space. Successive versions
introduced a more detailed structure of scalar fields, which we outline
here, and later brought the compactification to a sphere (which avoids
the problem of gauge bosons localization). This choice of compactification
is clearly not unique, though representative. In particular, warped
compactifications, like those considered in ref.\ \cite{Massimo}, lead to
a similar pattern of masses and mixings.

The basic set-up consists of a vortex structure (which acts as a kind of
background). It is created by a scalar field $\Phi$, charged under a
gauged $U(1)_{\rm g}$, to which is associated the gauge field\footnote{Our
notations coincide with those used in Refs.~\cite{LT, FLNT_sphere}. In
particular, 6-D coordinates are labeled by capital Latin
indices $A,B=0,\ldots,5$. Four-dimensional coordinates are labeled by
Greek indices, $\mu ,\nu =0,\ldots,3$. The signature is mostly negative.}
$A^M$. Fermions are introduced as one 6-D fermion field for
each charge of fermions (that means one $U$ field, which will later
generate the right-handed (4-D) up, charmed and top quarks, and so on for
$D$ down-type quark, $Q$ quark doublet, $E$ lepton singlet and $L$ lepton
doublet field). A singlet fermion $N$ is also included. Here we have a
choice: if we wish to deal with 4-D Dirac neutrinos, light right-handed
neutrinos are needed, and $N$ must couple to the vortex structure. While
this choice is perfectly possible, we take here the alternate view, and
seek a 4-D Majorana solution for the physical neutrinos. The $N$ field
will remain a ``bulk'' field in this paper, but will provide the Majorana
masses. These fermion fields are seen as 8-component fermions in 6-D, but
their couplings can be distinguished in terms of 6-D ``chirality'' -- to
avoid later confusion we will call it $\Gamma_7$ parity, and the $A^M$
fields couples  unequally to the ``$+$'' and ``$-$'' $\Gamma_7$ parity
components of the fermion fields.

A scalar doublet field $H$ provides the usual Brout-Englert-Higgs (BEH)
electroweak symmetry breaking, and also will contribute to the fermion
masses. More fields are needed to generate realistic masses (in
particular, to generate the off-diagonal couplings leading to mixing), and
we could choose to introduce them as needed (for instance, an extra scalar
doublet).  Since renormalizability is anyway lost in 6-D, we choose
instead to minimize the number of degrees of freedom by factorizing the
interaction. Thus, a single $SU(2)\times U(1)$ singlet, $X$ is introduced.
In the same way, since we will need a winding-3 vortex, we could request
the $\Phi$ field to exhibit such a solution, but in practice, we prefer to
keep $\Phi$ winding to 1, and introduce a $\Phi^3$ couplings to the
fermions. Of course, the opposite option could be taken, at the cost of
multiplying the fields.

Even with this restricted set of fields, various assignations of the
$U(1)_{\rm g}$ vortex charge are possible, and depend in particular on the
choice of flat or spherical geometry. Therefore, we are dealing with a
class of models, of which we select the minimal one for further
quantitative development. The present model is determined by the  charge
assignments of the fields  listed in table~\ref{tab:charges}.
\begin{table}[h!]
\centering
\begin{tabular}{|c|c|c|c|c|c|}
\hline\hline Field& Notation & $U(1)_g$ & $U(1)_Y$ & $SU(2)_w$    &
$SU(3)_c$ \\
\hline \hline vortex scalar&$\Phi$ & $+1$     & $0$      & $\mathbf{1}$
& $\mathbf{1}$\\
\hline BEH boson&$H$ & $0$      & $+1/2$   & $\mathbf{2}$ &
$\mathbf{1}$\\
\hline auxiliary scalar&$X$    & $+1$     & $0$      & $\mathbf{1}$ &
$\mathbf{1}$\\
\hline \hline quark $SU(2)_{\rm W}$ doublet&$(Q_{+},Q_{-})$ & $(3,0)$ &
$1/6$ & $\mathbf{2}$ & $\mathbf{3}$\\
\hline up-type quark $SU(2)_{\rm W}$ singlet&$(U_{+},U_{-})$    &
$(0,3)$ & $2/3$ & $\mathbf{1}$ & $\mathbf{3}$\\
\hline down-type quark $SU(2)_{\rm W}$ singlet&$(D_{+},D_{-})$    &
$(0,3)$  & $-1/3$ & $\mathbf{1}$ & $\mathbf{3}$\\
\hline lepton $SU(2)_{\rm W}$ doublet&$(L_{+},L_{-})$    & $(3,0)$  &
$-1/2$   & $\mathbf{2}$ & $\mathbf{1}$\\
\hline charged lepton $SU(2)_{\rm W}$ singlet&$(E_{+},E_{-})$    &
$(0,3)$ & $-1$     & $\mathbf{1}$ & $\mathbf{1}$\\
\hline sterile neutrino singlet&$(N_{+},N_{-})$    & $(0,0)$  & $0$ &
$\mathbf{1}$ & $\mathbf{1}$\\
\hline\hline
\end{tabular}
\caption{
\label{tab:charges}
Charge assignments of the fields under the gauge
groups of the SM and under $U(1)_{\rm g}$. For fermions, the two numbers
in parentheses correspond to the charges of the components with positive
and negative values of the $\Gamma_{7}$ parity, denoted everywhere by
``$+$'' and ``$-$'' indices, correspondingly.}
\end{table}
We note that the choice of charges in the scalar sector is slightly
different with respect to our previous work. The possibility of such
variants is discussed in section~\ref{sec:concl}.

\section{Scalar parameters}
\label{sec:scalar}

\subsection{Scalar profiles}
\label{sec:scalar-profiles}
In a first step, we adjust (within the above charge assignment) the model
parameters to generate the desired scale for electroweak symmetry breaking
(we anticipate here somewhat to the fact the KK scale will be bounded from
below by the absence of flavour-changing neutral currents). This involves
fixing the potential and  couplings of $\Phi$ , $H$ (and also the extra
field $X$). The two extra dimensions are compactified on a sphere of
radius $R$, with the spherical coordinates $\theta, \phi$.

Let us suspend for the moment the field $X$ which will be required to
generate fermion mixings. We will switch it on back in the next
subsection. As announced, the BEH field $H$ acquires a non-constant
profile in transverse dimensions due to its interaction with the scalar
$\Phi $. The role of the $H$ field in the model requires it to transform
under the SM electroweak gauge group $SU(2)_{\rm W}\times U(1)_{\rm Y}$ as
usual. In the SM, the electroweak symmetry is broken by the BEH-field
vacuum expectation value provided by the potential term $\sim
\left(|H|^2-\mu^2\right)^2$. In our case, the non-constant $H$ profile is
obtained with the help of the coupling between $\Phi$ and $H$, so the
complete Lagrangian for them reads \cite{LN_Higgs, LN_Higgs_Yad}:
\begin{equation}
\mathcal{L}_{\Phi+H}= R^2 \sin\theta
\left(\left|D_A\Phi\right|^2+\left|D_A H\right|^2
-\frac{\lambda}{2}\left(\left|\Phi\right|^2-v^2\right)^2
-\frac{\kappa}{2}\left(\left|H\right|^2-\mu^2\right)^2
-h^2\left|H\right|^2\left|\Phi\right|^2\right).
\label{LagPhiandH}
\end{equation}

Consider the last three terms of this expression. The coupling
$h^2\left|H\right|^2\left|\Phi\right|^2$ enforces ``orthogonalization'' of
the two scalar fields in the sense that the larger value of one
corresponds to a reduced value of the other. The topology of the vortex
determines the global profile in which $H(\theta)$
goes to zero at sufficiently large $\theta$. At the origin, where $\Phi\to
 0$, the last term loses its importance while the usual $\sim
\left(|H|^2-\mu^2\right)^2$ term enforces a nonzero value of $H$. For
sufficiently large $h^{2}$, this value gets smaller and the entire
solution approaches the trivial one which is phenomenologically
inappropriate. In the opposite case of very small $h^{2}$, the solution
for $H^{2}$ approaches its vacuum value $\mu^{2}$. This construction is
analogous to that of Witten \cite{Witten-SCstrings} where a similar
problem has been solved in four dimensions.

To find a nonsingular solution, one imposes the following boundary
conditions,
$$
H'(0)=0, \ \ H(\pi)=0.
$$
These conditions agree with the
physical picture suggested above. Note that the condition $H'(0)=0$ does
not fix the $H$ value at the origin: instead, this value is determined by
parameters of the potential. The latter are constrained by the
requirement that the solution should reproduce the proper SM BEH
vev~\cite{FLNT_flavour},
\begin{equation}
\frac{V_{SM} ^2}{2}=2\pi R^2\int_0^{\pi}\mathrm{d}\theta \sin\theta
H^2(\theta),
\label{normH}
\end{equation}
where $V_{SM}\approx 246$~GeV fixes the energy scale of the electroweak
symmetry breaking and determines, e.g., the W-boson mass.

The orthogonalization effect ensures that the characteristic extension
$\theta_{H}$ of the profile of the field $H$ is of the same order as
$\theta_{\Phi}$, the characteristic extension of $\Phi$. In practice, we
always have $\theta_{H}\lesssim \theta_{\Phi}\ll 1$ which allows to
simplify the normalization integral,
$$
\frac{V_{SM}^2}{2}\approx \pi
R^2\theta_{\Phi}^2 H^2(0),
$$
where we approximated $\sin\theta\approx
\theta$, put $H(\theta)\approx H(0)$ on the interval $[0,\theta_{\Phi}]$
and $H(\theta)\approx 0$ outside it. We further define
\begin{equation}
\Lambda_{\rm V}\equiv 1/(R\theta_{\Phi}),
\label{Eq:LambdaV}
\end{equation}
a parameter with the dimension
of energy which sets up the vortex scale. We obtain
$$
H(0)\approx
\left(\frac{0.1~\text{TeV}}{\Lambda_{\rm V}}\right)\Lambda_{\rm V}^2 .
$$
As we will see
below (section~\ref{sec:experiment}), phenomenological constraints from
the absence of rare processes require $\Lambda_{\rm V}\sim 10^4$~TeV, which
implies $H(0)\approx 10^{-5}\Lambda_{\rm V}^2$. One may note that this fine tuning
is a weak point of the model; however, it is a reflection of the
unexplained adjustment of parameters required to obtain the electroweak
scale in the SM. It is tempting to argue for a 6-D dynamical reason for
this tuning, cf. \cite{306, 307}. We stress that this is the only critical
fine-tuning in the model (as in grand unified theories).

By a careful study of the solution for the scalar fields, together with
the normalization requirement, eq.~(\ref{normH}), one may obtain an
interesting approximate relation, whose technical derivation is beyond the
scope of the present work,
\[
m_{\rm H}\simeq \kappa \mu \times 70 ~{\rm GeV},
\]
for the 4-D mass of the observable scalar-boson mode. One may note that
the dimensionless combination $\kappa \mu$ plays the role of a coupling
constant in the 6-D model and therefore should not be too large, say
$\kappa \mu \lesssim 4\pi$, to keep perturbativity. Therefore, weak
coupling implies a relatively light BEH boson in this model
~\cite{LN_Higgs, LN_Higgs_Yad}. It is interesting to note that this single
tuning of the 6-dimensional solution results in effectively decoupling the
EW breaking sector from the $1/R$ scale, which makes it almost
indistinguishable from a pure 4-D version of the SM.

The numerical search for the scalar solution is  technically  very
difficult because the trivial solution is very close to the interesting
one, so the standard procedure often finds the trivial, though unstable,
solution. Theoretical considerations have suggested~\cite{LN_Higgs,
LN_Higgs_Yad}, however, a way to find the stable solution which will be
used below.

\subsection{The $X$ field and the pseudo-Goldstone mode}
\label{sec:Goldstone}
As explained in our previous works~\cite{LT, FLT},
the model with two scalar fields, $\Phi$ and $H$, possesses an exact
symmetry, the family number, which is related to the geometry of the extra
dimensions and counts the winding number of the six-dimensional fields.
While the proper hierarchical masses of quarks and charged leptons may be
generated in this way, the CKM mixings are forbidden by this symmetry.
This is achieved by introducing the field $X$, cf.\
table~\ref{tab:charges}. The bosonic Lagrangian should be supplemented
with
\[
\mathcal{L}_{X}= R^2 \sin\theta \left(\left|D_A X\right|^2-
\frac{\rho}{2}\left(\left|X\right|^2-v^2\right)^2-
\eta^2\left|X\right|^2\left|\Phi\right|^2\right).
\label{LagX}
\]
In our basic formulation the scalar potential for $X$ and $\Phi$ is
insensitive to their relative phase, and thus possesses two global $U(1)$
symmetries which correspond to independent phase rotations of $X$ and
$\Phi$. Both symmetries are broken by nonzero scalar profiles we require.
This leads to two Goldstone bosons of which one combination is absorbed by
$A^M$ to give it mass while the other, which we will call $Y$  is left
massless at this stage.

Would such a (pseudo-) Goldstone boson be phenomenologically excluded? Its
couplings to light particles are determined by the Lagrangian, and depend
upon the evaluation of the corresponding overlaps; in general they are
suppressed by a factor of order $\langle H \rangle /\langle X \rangle$
with respect to the Standard Model Scalar. The charge assignment we have
chosen, table~\ref{tab:charges}, guarantees that these couplings are
furthermore off-diagonal in flavour (the key point here is the charge of
$H$ under $U(1)_{\rm g}$ which we selected in this paper expressly for
this reason). However, even such reduced couplings, which effectively
suppress virtual processes are dangerous if the pseudo-goldstone particle
is (nearly) massless, a typical test being $\mu \rightarrow e  Y$ decay.

In fact, we may notice that there are no anomalies associated to the $Y$
current, hence no axion-like behaviour, but on the other hand, the full
Lagrangian violates the associated symmetry at the level of the Yukawa
couplings, see table~\ref{tab:fermion-par}. Mass contributions (determined
by the high scale and the Yukawa couplings) would thus be generated at
loop level, which is enough to exclude production of $Y$ in low-energy
processes. Its coupling suppression then brings the expected limits in a
line with those from KK gauge-mediated flavour-changing currents.
Note in passing that there is no ``lowest stable KK mode'' in
this model, which could serve as a dark matter candidate. As in the
Standard Model, dark matter should be added explicitly, but is not
considered in the present publication.

We could of course introduce {\it ab initio} a coupling between the phases
of $X$ and $\Phi $, like the one induced by  these loop effects; the
difficulty then is in resolving coupled differential equations in the 2-D
space; we have preferred to keep this mass generation as a later
perturbation of the ``scalar background'' due to the fermionic fields.

\subsection{Fitting the scalar-boson mass}
\label{sec:fit-scalar}
The approach to numerical fitting of the scalar
parameters, quite nontrivial by itself, was outlined in
ref.~\cite{LN_Higgs, LN_Higgs_Yad}. One should take care of two principal
constraints whose origin lays outside of the scalar sector. One is the
relation between characteristic extensions $\theta_{\Phi}$ and
$\theta_{A}$ of the, respectively, scalar and gauge fields forming the
vortex,
$$
\sigma=\theta_\Phi / \theta_A \ll 1,
$$
which
ensures~\cite{FLNT_sphere} that the fermions may acquire hierarchical
masses, and another is
\begin{equation}
1/R \gtrsim 50~\mbox{TeV},
\label{Eq:50-TeV}
\end{equation}
as required by the
absence of dangerous flavour-changing processes, see
ref.~\cite{FLNT_flavour} and section~\ref{sec:experiment}. In practice,
the field $X$ is treated as a perturbation so the first step is to obtain
the correct scalar mass and effective vacuum expectation value,
eq.~(\ref{normH}), by solving the problem for $A$, $\Phi$ and $H$.
Skipping details, we obtain the parameters of the bosonic Lagrangian
listed in table~\ref{tab:scalar-param}.
\begin{table}
\centering
\begin{tabular}{|c|l|}
\hline\hline $R$       & $77.0 \ \Lambda_{\rm V}^{-1}$   \\
\hline $e$  & $0.01 \ \Lambda_{\rm V}^{-1}$   \\
\hline $v$  & $1.00 \ \Lambda_{\rm V}^2$ \\
\hline $\lambda$ & $1.00 \ \Lambda_{\rm V}^{-2}$   \\
\hline $\mu$ & $0.50 \ \Lambda_{\rm V}^2$ \\
\hline $\kappa$
& $4.22\ \Lambda_{\rm V}^{-2}$ \\
\hline $h^2$     & $1.64\ \Lambda_{\rm V}^{-2}$   \\
\hline $v$ & $1.50 \ \Lambda_{\rm V}^2$   \\
\hline $\rho$    & $0.50 \ \Lambda_{\rm V}^{-2}$   \\
\hline $\eta^2$  & $1.20 \ \Lambda_{\rm V}^{-2}$   \\
\hline\hline
\end{tabular}
\caption{
\label{tab:scalar-param}
Parameters of the compactification and of the
scalar sector. We have rounded up the value of $\kappa$ which is tuned to
a higher precision, as discussed in the text, to ensure the low value of
the weak scale compared to the vortex size.}
\end{table}
With the overall scale $\Lambda_{\rm V} \simeq 7700$~TeV we reproduce both
parameters of the Standard Model scalar sector, $m_H \simeq 125$~GeV and
$V_{SM}\simeq 246$~GeV. We note here that the number of parameters in the
scalar sector of our model exceeds that of the SM; this will be
compensated by a highly predictive fermionic sector which we consider in
the next section.

\section{Fermion parameters}
\label{sec:fermion}
\subsection{Selection rules and hierarchies for
charged fermions and neutrinos}
\label{sec:hierarchies}
The topology of the background bosonic field
provides a certain number of fermionic chiral zero modes thanks to the
index theorem, as explained e.g.\ in \cite{JackiwRossi} for (2+1)
dimensions and in \cite{Witten-SCstrings} for (3+1) dimensions. In our
case, the coupling is selected to have the topological number three, so
three zero modes appear in the effective (3+1)-dimensional theory for each
(5+1)-dimensional fermion. These modes are linearly independent which in
our case guarantees that they have different winding numbers, ${\rm
e}^{ik\phi}$ with $k=0,1,2$, in the two extra dimensions. The different
windings imply in turn the different behaviour, $\sim \theta^{k}$, at the
origin. Upon the electroweak symmetry breaking, that is in the presence of
$H\ne0$, these zero modes are lifted and the hierarchical mass matrix
emerges \cite{LT, FLT, FLNT_sphere}, with the values determined by the
overlaps between the scalar and fermion profiles, multiplied by the Yukawa
couplings. Our model includes the following coupling of fermions to the
vortex:
\begin{align}
& g^{\Psi} \Phi^k \bar{\Psi}_+\Psi_- + h.c. \quad \text{will provide the
left-handed 4-D species},\nonumber\\
& g^{\Psi} \Phi^k \bar{\Psi}_-\Psi_+ + h.c. \quad \text{will provide
the right-handed 4-D species},\nonumber
\end{align}
where $\Psi=(\Psi_+,\Psi_-)$ stands for $Q$, $U$, $D$, $L$ and $E$ 6-D
fermions and $k$ gives the number of 4-D zero modes for each of these
 species. The charges for fermions $(Q(\Psi_+),Q(\Psi_-))$ under the
  vortex $U(1)_g$ group must be chosen such that $|Q(\Psi_+)-Q(\Psi_-)|=k
  Q(\Phi)$. In our working example, $k=3$ and the assignments are given in
  the third column of table~\ref{tab:charges}. It's worth mentioning that
 the 4-D chirality is determined by the sign of $Q(\Psi_+)-Q(\Psi_-)$.
The zero modes are chiral also in the six-dimensional sense, which means
that in a particular (``chiral'') representation of 6-D Dirac matrices,
see~\cite{LT}, certain components of 8-component spinors are zero. In
particular, for the quarks we have the following structure of the zero
modes,
\begin{equation}
Q_{k}= \left(
\begin{tabular}{c}
0\\
${\rm e}^{-i\phi(k-3)}\dots$\\
${\rm e}^{-i\phi(k-1)}\dots$\\
0
\end{tabular}
\right) , \quad U_{k}= \left(
\begin{tabular}{c}
${\rm e}^{-i\phi(k-1)}\dots$\\
0\\
0\\
${\rm e}^{-i\phi(k-3)}\dots$
\end{tabular}
\right) ,
\label{Eq:A1*}
\end{equation}
$k=1,2,3$, and similarly for the $SU(2)_{\rm W}$-singlet (right-handed)
$D$ quarks (in 6-D, spinors are 8-component and each element here
represents a bispinor). The couplings which lift the quark zero modes read
as
\[
Y_{d} H\bar{Q}_+ D_- + Y_{u} \tilde{H}\bar{Q}_+ U_-,
\]
where $\tilde H$ is obtained from $H$ by the action of the
antisymmetric tensor $\epsilon$ in the $SU(2)$ space, $\tilde H_{i} =
\epsilon_{ij} H_{j}$. These interaction terms result in the effective
four-dimensional mass matrix
\[
m_{kl}=\int \! d\phi d\theta H \bar Q_{k} D_{l}
\]
and the structure (\ref{Eq:A1*}) results in,
\begin{equation}
m_{kl}\propto \delta_{kl}\cdot\sigma ^{2(3-k)},
\label{Eq/Pg11/1:Update6Dv5}
\end{equation}
where the selection rule $m_{kl}\sim\delta _{kl}$ comes from the $\phi$
integration, while the $\theta $ integration gives rise to hierarchical
dependence on $\sigma $. The selection rule is of course a manifestation
of the family-number symmetry; it leads to a diagonal but (because of
overlaps in $\theta$ integration) strongly hierarchical mass matrix. The
off-diagonal elements, corresponding to the CKM mixing, can be populated
by introducing an additional field $X$ which allows to write interactions
with $\Phi$ whose profile breaks the 6-D rotational symmetry, and
therefore the family-number symmetry, spontaneously. The hierarchy of the
mass matrix elements is generated by a small parameter
$\sigma=\theta_{\Phi}/\theta_{A}$.

For charged leptons, the interaction which produces diagonal masses is
given by the similar term
\[
Y_{e} H \bar{L}_+ E_-
\]
which results in a similar hierarchical mass structure due to similar
wave-function profiles. It would be also the case for neutrinos if they
had a Dirac mass (see~\cite{FLT_neutrino}); however, for the effective 4-D
Majorana masses, the situation is quite different~\cite{FLL_neutrino}. The
corresponding interaction is given by
\[
Y_{n} \tilde{H} \bar{L}_-N_+
\]
for the $N$ field. In the present approach (where we seek light
Majorana neutrinos in 4-D), the $N$ field remains a ``bulk'' field (no
light modes), and does not have coupling to the vortex. The wave functions
of its modes should be normalized and their absolute value is therefore
suppressed, thus diminishing the overlaps. There exists a 6-D mass term
for the $N$ field,
\begin{equation}
\label{Maj}
\frac{M}{2}\bar{N^{c}} N +\mbox{h.c.}
\end{equation}
Of course, no Majorana fermions exist in 6-D. The above term
(eq.~(\ref{Maj})), while evocative of a Majorana mass term, in fact
couples $N_+$ to $N_-$. Still leptonic number violation results from the
Yukawa couplings, and eq.~(\ref{Maj}) leads to light 4-D Majorana masses
suppressed by the see-saw mechanism for the physical
neutrinos~\cite{FLL_neutrino}. While the zero modes of the field $L$ have
a Lorentz structure similar to eq.~(\ref{Eq:A1*}), the structure of the
Majorana interaction results in a different selection rule for the masses,
\begin{equation}
m_{kl} \propto \delta_{4,k+l}\cdot\sigma ^{3-|k-l|/2},
\label{Eq/Pg12/1:Update6Dv5}
\end{equation}
which therefore provides a very different (anti-diagonal) hierarchical
structure, which is in the end surprisingly similar to that which emerges
from the experimental data. The overall suppression of neutrino masses is
achieved by the usual see-saw mechanism, where the large mass for the
singlet field $N$, which is not localized on the brane, appears as a
parameter of the model which is naturally of order of the compactification
scale. This structure is the direct cause for the most striking feature of
the model, large lepton mixing \cite{FLL_neutrino}.

It is important to note that $\sigma $ in eq.~(\ref{Eq/Pg12/1:Update6Dv5})
is the same as in eq.~(\ref{Eq/Pg11/1:Update6Dv5}), so that both
hierarchies are governed by the same parameter. It determines also the
value of the neutrino mixing angle $\theta _{13}\sim\sigma $.

\subsection{Fitting the masses and mixings}
\label{sec:masses}
Therefore, the mechanism we advocate incorporates two
very different hierarchies within a single framework. However, the
hierarchies of masses of charged fermions differ (e.g. for the quarks, the
mass ratios are much more severe for the ``up'' sector than for the
``down'' sector). Various strategies are possible to get realistic masses
and mixing.  This can be fixed, either through mixing terms, or by
choosing the initial parameters to optimize the diagonal ratios, hence
different profiles for the $U$ and $D$ fermions. We present here a
benchmark choice of parameters which allow for a nice fit of the
experimental masses and mixings.

Since the fermion masses experience renormalization-group running
(especially important for strongly-interacting quarks), at which energy
 scale should we fit the masses\footnote{The first one who asked this
question was M.~Voloshin.}? Clearly, it is hard to give a precise answer.
In our mechanism, like in the SM, fermion masses are associated with the
electroweak breaking, so it is natural to calculate their values at this
scale. We therefore use the values at $M_{Z}$, the scale which is commonly
used. One shall note however that for light quarks, the mass running in
the strong-coupling regime (momenta $\lesssim \Lambda_{\rm QCD}$) is
uncertain. As a result, in our fit, we concentrate on obtaining more
precise values for masses of heavier quarks and all leptons, as well as
for mixing angles (whose running is negligible \cite{no-angle-run}),
relaxing to some extent uncertainties in masses of $u$ and $d$ quarks.

Parameters of the model in this particular fit are listed in
table~\ref{tab:fermion-par}.
\begin{table}
\centering
\begin{tabular}{|l|l|}
\hline\hline \multicolumn{2}{|c|}{Fermion-vortex couplings}\\
\hline $g_L^Q$ & $0.020\ \ \Lambda_{\rm V}^{-5}$   \\
\hline $g_R^U$ & $0.005\ \ \Lambda_{\rm V}^{-5}$   \\
\hline $g_R^D$ & $0.063\ \ \Lambda_{\rm V}^{-5}$ \\
\hline $g_L^L$ & $0.013\ \ \Lambda_{\rm V}^{-5}$   \\
\hline $g_R^E$ & $0.013\ \ \Lambda_{\rm V}^{-5}$   \\
\hline\hline \multicolumn{2}{|c|}{``Yukawa'' fermion-scalars
couplings}\\
\hline $H\bar{Q}_+ D_-$ & $7.00\ \ \Lambda_{\rm V}^{-1}$ \\
\hline $H X^* \Phi \bar{Q}_+ D_-$ & $14.0\ \ \Lambda_{\rm V}^{-5}$   \\
\hline \hline $\tilde{H}\bar{Q}_+ U_-$ & $850\ \
\Lambda_{\rm V}^{-1}$ \\
\hline $\tilde{H} X^* \Phi\bar{Q}_+ U_-$ & $255\ \
\Lambda_{\rm V}^{-5}$ \\
\hline \hline $H X^*\Phi\bar{L}_+ E_-$ & $56\ \ \Lambda_{\rm V}^{-5}$ \\
\hline $H X \Phi^* \bar{L}_+ E_-$ & $-160\ \ \Lambda_{\rm V}^{-5}$ \\
\hline \hline $\tilde{H} \bar{L}_-N_+$ & $0.165\ \
\Lambda_{\rm V}^{-1}$   \\
\hline $\tilde{H} X\Phi^* \bar{L}_-N_+$ & $(0.91+1.04i)\cdot 10^{-2}\ \
\Lambda_{\rm V}^{-5}$ \\
\hline $\tilde{H} (X^*)^4 \Phi \bar{L}_+N_-$   & $1.00\ \
\Lambda_{\rm V}^{-11}$   \\
\hline\hline
\end{tabular}
\caption{
\label{tab:fermion-par}
Benchmark parameters for the 6-D fermionic sector,
given in terms of the overall scale $\Lambda_{\rm V}$ which in our case is
$\approx 7700$~TeV. It is worth noting that, while necessary for mass
generation, the last coupling is fully arbitrary and can always be put
equal to $1$ by redefinition of other Yukawa's couplings in the neutrino
sector. Then it can be disregarded in the counting of free parameters of
the model.}
\end{table}
The values of the mass-matrix observables for this fit, together with
their experimental values \cite{PDG, massesAtMZ, NuFit} (calculated at
$M_{Z}$ when applicable), are given in table~\ref{tab:masses}.

\begin{table}
\centering
\begin{tabular}{|c|c|c|}
\hline\hline Parameter & Fitted value & Experimental value\\
\hline \multicolumn{3}{|c|}{The scalar-boson mass} \\
\hline $m_H$   & 125~GeV & $125.5\pm0.2 \mbox{(stat.)} \pm 0.6
\mbox{(syst.)}$ \cite{boson:ATLAS:properties} \\
& & $125.7\pm 0.3 \mbox{(stat.)} \pm 0.3 \mbox{(syst.)}$
\cite{boson:CMS:properties}\\
\hline \multicolumn{3}{|c|}{Quark masses at Z scale}  \\
\hline $m_{d}$ &0.01~GeV &  $(0.00282 \pm 0.00048)$~GeV \\
$m_{s}$ &0.051~GeV  & $(0.057^{+0.018}_{-0.012})$~GeV  \\
$m_{b}$ &2.86~GeV  & $2.86^{+0.16}_{-0.06}$~GeV \\
$m_{u}$ &0.023~GeV & $0.00138^{+0.00042}_{-0.00041}$~GeV  \\
$m_{c}$ &0.72~GeV  & $0.638^{+0.043}_{-0.084}$~GeV \\
$m_{t}$ &172~GeV & $172.1\pm 1.2$~GeV\\
\hline \multicolumn{3}{|c|}{Quark mixing matrix} \\
\hline $\left\vert U_{\text{CKM}}\right\vert$ & $
\begin{pmatrix}
0.979 & 0.207 & 0.0015\\
0.206 & 0.9730 & 0.046\\
0.011 & 0.049 & 0.999
\end{pmatrix}
$ & $
\begin{pmatrix}
0.97427\pm 0.00015 & 0.22534 \pm 0.00065 & 0.00351
^{+0.00015}_{-0.00014}\\
0.22520\pm 0.00065 & 0.97344\pm 0.00016 & 0.0412^{+0.0011}_{-0.0005}\\
0.00867^{+0.00029}_{-0.00031} & 0.0404^{+0.0011}_{-0.0005} &
0.999146^{+0.000021}_{-0.000046}
\end{pmatrix}
$ \\
\hline \multicolumn{3}{|c|}{Charged-lepton masses}  \\
\hline $m_{e}$ & 0.00061~GeV  & 0.0004866~GeV  \\
$m_{\mu}$ &0.089~GeV & 0.1027~GeV \\
$m_{\tau}$ &1.74~GeV  &  1.746~GeV \\
\hline \multicolumn{3}{|c|}{Neutrino masses}  \\
\hline $m_{1}$ & $5.46\cdot 10^{-2}$~eV & -- \\
$m_{2}$ & $5.53\cdot 10^{-2}$~eV & -- \\
$m_{3}$ & $4.17\cdot 10^{-5}$~eV & -- \\
$\Delta m_{21}^2$ & $7.96\cdot 10^{-5}~\text{eV}^2$ & $(7.50\pm 0.185)
\cdot 10^{-5}~\text{eV}^2$\\
$\Delta m_{13}^2$ & $2.98\cdot 10^{-3}~\text{eV}^2$ &
$(2.47^{+0.069}_{-0.067})\cdot 10^{-3}~\text{eV}^2$\\
\hline \multicolumn{3}{|c|}{Lepton mixing matrix} \\
\hline $\left\vert U_{\text{PMNS}}\right\vert$ & $
\begin{pmatrix}
0.76 & 0.63 & 0.13\\
0.39 & 0.58 & 0.72\\
0.52 & 0.52 & 0.68
\end{pmatrix}
$ & $\simeq
\begin{pmatrix}
0.795 - 0.846 &   0.513 - 0.585 &   0.126 - 0.178\\
0.205 - 0.543 & 0.416 - 0.730 &   0.579 - 0.808\\
0.215 - 0.548 & 0.409 - 0.725 & 0.567 - 0.800
\end{pmatrix}
$ \\
$\left\langle m_{
\beta\beta}
\right\rangle$ & 0.013~eV & $\lesssim 0.3$~eV\cite{BilenkiBetaBeta}\\
J & 0.019 & $\lesssim 0.036$\\
$\theta_{12}$ & $39.7^{\circ}$ & $\simeq \left(31.09^{\circ} -
35.89^{\circ} \right)$\\
$\theta_{23}$ & $46.5^{\circ}$ & $\simeq \left(35.8^{\circ} -
54.8^{\circ} \right)$\\
$\theta_{13}$ & $7.2^{\circ}$ & $\simeq \left(7.19^{\circ} -
9.96^{\circ} \right)$\\
\hline\hline
\end{tabular}
\caption{
\label{tab:masses}
Fitted versus experimental values of the mass
parameters. If other is not explicitly stated, we use the values from
\cite{PDG} for CKM, ref.~\cite{massesAtMZ} for running charged-fermion
masses at $M_{Z}$ and ref.~\cite{NuFit} for neutrino parameters. For the
inverted hierarchy scenario, which is relevant for our model,
0.013~eV$\lesssim \left\langle m_{
\beta\beta}
\right\rangle \lesssim$0.05~eV. Experimental values of charged-lepton
masses are known to a better precision than quoted.}
\end{table}
Variations of the model, namely relaxing the quark-lepton universality of
$U(1)_{g}$ charges (which we prefer to avoid), may allow for a slightly
better fit.

Note that in the SM, redefinition of fields may always be used to remove
all unphysical phases from the mass matrix, so that only one physical
$CP$-violating phase remains in CKM. It is also true here; however, we
restricted ourselves for simplicity (in a first approximation)  to real
couplings and succeeded in obtaining a reasonable set  of quark masses and
mixing angles. We clearly could not (and did not) obtain the
$CP$-violating phase in this way.  We can introduce it by means of a free
complex phase parameter in the couplings but we cannot predict it. This
problem will be studied elsewhere, with ref.~\cite{CP} providing one of
possible frameworks.

The opposite happens in the leptonic sector. Here, we needed to use
complex couplings to obtain a reasonable fit to the experimental masses
and mixings. Therefore, the model presented here predicts non zero $CP$
violation in the neutrino sector whose measurable value may be
parametrized by the Jarlskog invariant (adapted for leptons), $J$, which
we also list in table~\ref{tab:masses}. Unlike for quarks, $CP$ violation
in neutrinos is a prediction of our model, at least in the present fit.

\section{Experimental tests}
\label{sec:experiment}

While the most impressive implications of the 6-D model is in the very
different nature of the neutrino sector, it is important to look for other
possible signals and tests. We have in previous papers suggested
experimental tests and constraints. Here, we give a brief review and
update on them.

The most constraining ones are related to the KK excitations of gauge
bosons, which carry ``winding'', i.e. family number. We typically get
neutral flavour-changing, but family number-conserving interactions, and a
strong bound on $( \varkappa R )^2$ is obtained, where $R$ is the
compactification scale, and $\varkappa$ is a parameter describing the
convolution of the overlap between wave functions of fermions from
different families and  the gauge-boson KK-mode profile. This suppresses
the coupling with respect to the canonical  (diagonal) gauge couplings. In
previous work~\cite{FLNT_flavour, B_mesons}, we have obtained modest
suppression factors $\varkappa$, leading to an expected mass scale for the
KK excitations, well beyond LHC reach, but still a challenge for
increasing precision low energy tests. While the previous calculations
were based on approximate wave functions, for this study, we performed a
detailed calculation of the overlaps, starting from our benchmark
parameters and numerical solutions for the spherical compactification. The
strongest constraint still comes from the $K_L^0\to \bar{\mu}e$ (or $\to
\mu\bar{e}$) decay and reads
\[
1/R \gtrsim 50~\mbox{TeV}.
\]
This is the constraint we used when tuning the scalar sector. The
prediction of any model with slightly broken family symmetry would be the
observation of $K\to\overline{\mu }e$ decay as well as $B^{0}\to
K^{0}\overline{\mu} e$ (or their charge conjugate) decay~\cite{B_mesons}
at a level less suppressed than either $\mu\to e\gamma$ or $\mu \to
e\overline{e}e$, or $\mu -  e$ conversion on  nuclei (the latter FCNC
processes indeed violate family number). \bigskip

As pointed out before~\cite{FLNT_LHC}, the models we are considering here
may be generic of similar ones with more general geometry (or simply a yet
unexplored range of parameters)\footnote{In the framework chosen in the
present paper, we would need to reduce strongly $1/R$ and the overlaps
$\varkappa$ accordingly. We have found that this possibility was strongly
restricted,  in particular by  the smallness of the  Standard Model Scalar
mass (125 GeV). Alternatives would at least imply modifications in the
scalar potential and/or in the geometry of the compactification. These
considerations lie outside the scope of the present work.}. In fact, the
above constraint on the KK scale rests entirely on the value of the
overlaps $\varkappa$. Models or fits providing lower $\varkappa$ values
would allow to reduce the mass of the $Z'$ and other recurrences, and
maybe bring them in the reach of accelerator searches. Such an analysis
can be conducted in a fairly model-independent way, since it directly
relates constraints from precision data to LHC searches. For this reason,
a search for possible signals is certainly in order already at LHC. Of
course, the price to pay for a low $\varkappa$ is also a reduced
production of the recurrences. The relevant processes are the production
of $\gamma'$, $Z'$ or $W'$ bosons  (the actual KK mode) which carry a
nonzero winding number and thus decays with family violation. The signal
is indeed quite striking ($\mu^+ e^-$ pairs exceeding $\mu^- e^+$ pairs,
for instance).

Here we  ignore the KK recurrences of the gluons. They are of course easy
to include, but the corresponding signals are difficult to disentangle (as
they need precise quark identification) while enjoying larger production
rates. Another point is that we refer to $Z'$, $\gamma'$, etc.; it should
be noted that, in first approximation (ignoring the scale of EW symmetry
breaking as compared to the compactification scale), those modes are
degenerate and should thus be added coherently. It has been argued (see
ref.~\cite{Cheng:2002iz}) that loop corrections might instead lift the
degeneracy into a $W_3'$-$B'$ mode rather than $Z'$-$\gamma'$ way. We
leave this question open, as we are rather wary of getting into
(non-renormalizable) loop calculations.

\bigskip

We have checked that the couplings of the scalar boson $H$ to vector
bosons and fermions are very similar to the SM, and therefore we do not
predict any particular signature here. This is true for tree-level
couplings, but for loop-induced couplings as well (like $\gamma \gamma$ or
gluon-gluon), to which the KK contributions appear to be negligible.
Note that the bound (\ref{Eq:50-TeV}) is much stronger than the ones
usually quoted to satisfy other precision tests, like the $S$ and $T$
parameters (even if the context is somewhat different here)
\cite{bounds}
and that they arise at the (safer from our point of view) tree level,
instead as through loops like e.g.\ in Ref.~\cite{BsGammaLoops}.

\bigskip

\begin{figure}
\centering \includegraphics[width=0.9\textwidth]{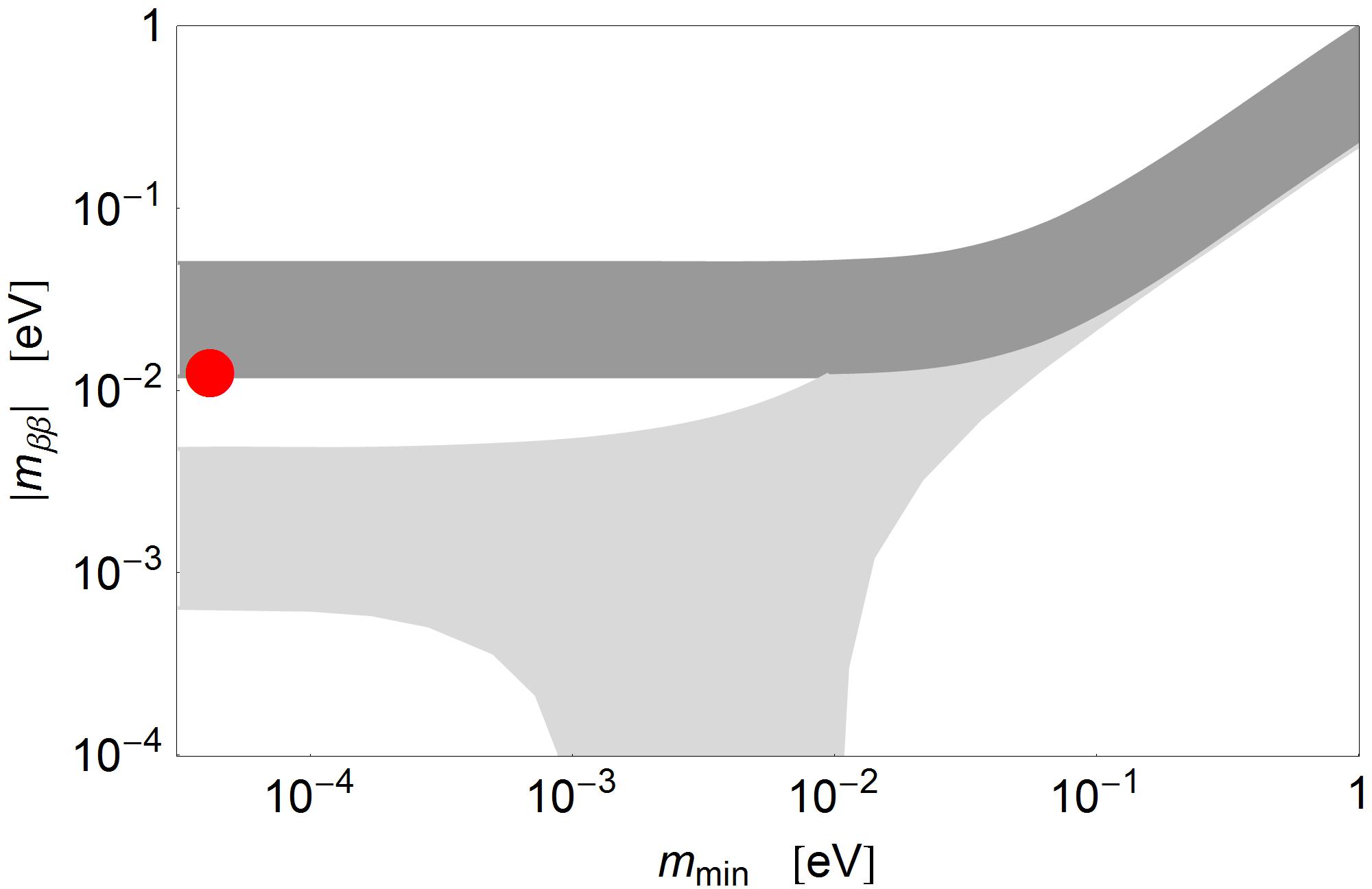}
\caption{
\label{fig:BetaBeta}
The effective Majorana mass $m_{
\beta\beta}
$ versus the minimal absolute neutrino mass. The 99\% allowed regions for
the normal hierarchy (light gray) and inverse hierarchy (dark
gray)~\cite{BilenkiBetaBeta} are shown together with the prediction of our
benchmark model (the thick dot).}
\end{figure}

Still, the most striking bunch of tests (not depending on
reaching the energy or sensitivity associated to the KK scale) remains in
the leptonic sector, where definite predictions are done. While large
mixings are a generic prediction of the approach, our benchmark fit
predicts inverted hierarchy, a particular value of $CP$ violation and the
smallest possible range of neutrinoless double beta decay compatible with
inverted hierarchy \cite{BilenkiBetaBeta, Rodejohann}, cf.\
section~\ref{sec:masses} (see Figure~\ref{fig:BetaBeta}).
This sensitivity will only be reached by the ``phase 3'' of the GERDA
experiment \cite{Gerda}.

\bigskip

A falsifying test for the present version of the model would be a
discovery of the fourth active neutrino. While the model is capable of
accommodating more than three families, its nice feature of the natural
explanation of large neutrino mixings within the same mechanism which
provides the quark mass hierarchy would fail for even number of families.
The case of five generations is beyond the scope of the present work.

\section{Conclusions}
\label{sec:concl}
To summarize, we review and update the six-dimensional
model of flavour which gives a natural and simultaneous explanation of two
very different mass-matrix hierarchies: one for charged fermions and
another one for neutrinos. Both hierarchies are governed by the same small
parameter. We formulate a benchmark model which gives a satisfactory fit
of all masses and mixings, given the latest measurements of the
scalar-boson mass and of neutrino mixing angles and taking advantage of
new, more precise, numerical solutions of the field equations. We give
distinctive experimental predictions which would allow to test the model
in future experiments. Some alternatives are still possible. For instance,
the charge assignments in table~\ref{tab:charges} could be modified, which
could change the perceived hierarchy of operators while  the full
parameter space has of course not been explored.

\section*{Acknowledgments} We are indebted to  D.~Gorbunov, E.~Nugaev and
M.~Tytgat for helpful discussions. M.L.\ and S.T.\ thank the Service de
Physique Th\'{e}orique at Universit\'{e} Libre de Bruxelles for kind
hospitality. This work is funded in part by IISN and by Belgian Science
Policy (IAP VII/37). This work was supported in part by the grant of the
President of the Russian Federation NS-5590.2012.2 (M.L.\ and S.T.); by
the Russian Foundation for Basic Research (grants 12-02-00653 (M.L.);
11-02-01528, 12-02-01203, 13-02-01311 and 13-02-01293 (S.T.)); by the RF
Ministry of Science and Education (agreements 8142 (S.T.) and
14.B37.21.0457 (M.L.\ and S.T.)) and by the ``Dynasty'' foundation (M.L.).


\begin{thebibliography}{38}

\bibitem{boson:ATLAS}
{\bf ATLAS} Collaboration, G.~Aad et~al., {\it {Observation of a new particle
  in the search for the Standard Model Higgs boson with the ATLAS detector at
  the LHC}},  {\em Phys.Lett.} {\bf B716} (2012) 1--29,
  [\href{http://xxx.lanl.gov/abs/1207.7214}{{\tt arXiv:1207.7214}}].

\bibitem{boson:CMS}
{\bf CMS} Collaboration, S.~Chatrchyan et~al., {\it {Observation of a new boson
  at a mass of 125 GeV with the CMS experiment at the LHC}},  {\em Phys.Lett.}
  {\bf B716} (2012) 30--61, [\href{http://xxx.lanl.gov/abs/1207.7235}{{\tt
  arXiv:1207.7235}}].

\bibitem{boson:ATLAS:properties}
{\bf ATLAS} Collaboration, B.~Mansoulie et~al., ``{Combination of results on
  the BEH Boson in ATLAS}.'' {Talk at the "XLVIIIth Rencontres de Moriond
  session devoted to Electroweak interactions and unified theories", La
  Thuile}, {March 2-9}, 2013.

\bibitem{boson:CMS:properties}
{\bf CMS} Collaboration, M.~Chen et~al., ``{Combination and interpretation of
  Scalar Boson search results from CMS}.'' {Talk at the "XLVIIIth Rencontres de
  Moriond session devoted to Electroweak interactions and unified theories", La
  Thuile}, {March 2-9}, 2013.

\bibitem{BroutEnglert}
F.~Englert and R.~Brout, {\it {Broken Symmetry and the Mass of Gauge Vector
  Mesons}},  {\em Phys.Rev.Lett.} {\bf 13} (1964) 321--323.

\bibitem{Higgs}
P.~W. Higgs, {\it {Broken Symmetries and the Masses of Gauge Bosons}},  {\em
  Phys.Rev.Lett.} {\bf 13} (1964) 508--509.

\bibitem{K2K}
{\bf T2K} Collaboration, K.~Abe et~al., {\it {Indication of Electron Neutrino
  Appearance from an Accelerator-produced Off-axis Muon Neutrino Beam}},  {\em
  Phys.Rev.Lett.} {\bf 107} (2011) 041801,
  [\href{http://xxx.lanl.gov/abs/1106.2822}{{\tt arXiv:1106.2822}}].

\bibitem{DayaBay}
{\bf DAYA-BAY} Collaboration, F.~An et~al., {\it {Observation of
  electron-antineutrino disappearance at Daya Bay}},  {\em Phys.Rev.Lett.} {\bf
  108} (2012) 171803, [\href{http://xxx.lanl.gov/abs/1203.1669}{{\tt
  arXiv:1203.1669}}].

\bibitem{LT}
M.~Libanov and S.~V. Troitsky, {\it {Three fermionic generations on a
  topological defect in extra dimensions}},  {\em Nucl.Phys.} {\bf B599} (2001)
  319--333, [\href{http://xxx.lanl.gov/abs/hep-ph/0011095}{{\tt
  hep-ph/0011095}}].

\bibitem{FLT}
J.~Frere, M.~Libanov, and S.~V. Troitsky, {\it {Three generations on a local
  vortex in extra dimensions}},  {\em Phys.Lett.} {\bf B512} (2001) 169--173,
  [\href{http://xxx.lanl.gov/abs/hep-ph/0012306}{{\tt hep-ph/0012306}}].

\bibitem{FLT_neutrino}
J.~Frere, M.~Libanov, and S.~V. Troitsky, {\it {Neutrino masses with a single
  generation in the bulk}},  {\em JHEP} {\bf 0111} (2001) 025,
  [\href{http://xxx.lanl.gov/abs/hep-ph/0110045}{{\tt hep-ph/0110045}}].

\bibitem{LN_fit}
M.~V. Libanov and E.~Y. Nougaev, {\it {Towards the realistic fermion masses
  with a single family in extra dimensions}},  {\em JHEP} {\bf 0204} (2002)
  055, [\href{http://xxx.lanl.gov/abs/hep-ph/0201162}{{\tt hep-ph/0201162}}].

\bibitem{LN_fit_t}
M.~Libanov and E.~Y. Nugaev, {\it {Hierarchical fermionic mass pattern and
  large extra dimensions}},  {\em Surveys High Energ.Phys.} {\bf 17} (2002)
  165--171.

\bibitem{FLNT_sphere}
J.~Frere, M.~Libanov, E.~Nugaev, and S.~V. Troitsky, {\it {Fermions in the
  vortex background on a sphere}},  {\em JHEP} {\bf 0306} (2003) 009,
  [\href{http://xxx.lanl.gov/abs/hep-ph/0304117}{{\tt hep-ph/0304117}}].

\bibitem{FLNT_flavour}
J.~Frere, M.~Libanov, E.~Nugaev, and S.~V. Troitsky, {\it {Flavor violation
  with a single generation}},  {\em JHEP} {\bf 0403} (2004) 001,
  [\href{http://xxx.lanl.gov/abs/hep-ph/0309014}{{\tt hep-ph/0309014}}].

\bibitem{FLNT_LHC}
J.~Frere, M.~Libanov, E.~Y. Nugaev, and S.~V. Troitsky, {\it {Searching for
  family number conserving neutral gauge bosons from extra dimensions}},  {\em
  JETP Lett.} {\bf 79} (2004) 598--601.

\bibitem{LN_Higgs}
M.~Libanov and E.~Y. Nugaev, {\it {Properties of the Higgs particle in a model
  involving a single unified fermion generation}},  {\em Phys.Atom.Nucl.} {\bf
  70} (2007) 864--870.

\bibitem{LN_Higgs_Yad}
M.~Libanov and E.~Nugaev, {\it {Higgs boson with a single generation in the
  bulk}},  \href{http://xxx.lanl.gov/abs/hep-ph/0512223}{{\tt hep-ph/0512223}}.

\bibitem{FLL_neutrino}
J.-M. Frere, M.~Libanov, and F.-S. Ling, {\it {See-saw neutrino masses and
  large mixing angles in the vortex background on a sphere}},  {\em JHEP} {\bf
  1009} (2010) 081, [\href{http://xxx.lanl.gov/abs/1006.5196}{{\tt
  arXiv:1006.5196}}].

\bibitem{Libanov:2011zz}
M.~Libanov and F.~Ling, {\it {Why neutrinos are different?}},  {\em PoS} {\bf
  QFTHEP2011} (2011) 072.

\bibitem{Libanov:2011st}
M.~Libanov and F.-S. Ling, {\it {Flavour puzzle or Why neutrinos are
  different?}},  \href{http://xxx.lanl.gov/abs/1105.6035}{{\tt
  arXiv:1105.6035}}.

\bibitem{Massimo}
M.~Giovannini, H.~Meyer, and M.~E. Shaposhnikov, {\it {Warped compactification
  on Abelian vortex in six-dimensions}},  {\em Nucl.Phys.} {\bf B619} (2001)
  615--645, [\href{http://xxx.lanl.gov/abs/hep-th/0104118}{{\tt
  hep-th/0104118}}].

\bibitem{Witten-SCstrings}
E.~Witten, {\it {Superconducting Strings}},  {\em Nucl.Phys.} {\bf B249} (1985)
  557--592.

\bibitem{306}
W.~D. Goldberger and M.~B. Wise, {\it {Renormalization group flows for brane
  couplings}},  {\em Phys.Rev.} {\bf D65} (2002) 025011,
  [\href{http://xxx.lanl.gov/abs/hep-th/0104170}{{\tt hep-th/0104170}}].

\bibitem{307}
E.~Dudas, C.~Papineau, and V.~Rubakov, {\it {Flowing to four dimensions}},
  {\em JHEP} {\bf 0603} (2006) 085,
  [\href{http://xxx.lanl.gov/abs/hep-th/0512276}{{\tt hep-th/0512276}}].

\bibitem{JackiwRossi}
R.~Jackiw and P.~Rossi, {\it {Zero Modes of the Vortex - Fermion System}},
  {\em Nucl.Phys.} {\bf B190} (1981) 681.

\bibitem{no-angle-run}
E.~Ma and S.~Pakvasa, {\it {Variation of mixing angles and masses with
$Q^{2}$ in the standard six quark model}},  {\em Phys.Rev.} {\bf D20}
(1979) 2899.

\bibitem{PDG}
{\bf Particle Data Group} Collaboration, J.~Beringer et~al., {\it {Review of
  Particle Physics (RPP)}},  {\em Phys.Rev.} {\bf D86} (2012) 010001.

\bibitem{massesAtMZ}
Z.-z. Xing, H.~Zhang, and S.~Zhou, {\it {Impacts of the Higgs mass on vacuum
  stability, running fermion masses and two-body Higgs decays}},  {\em
  Phys.Rev.} {\bf D86} (2012) 013013,
  [\href{http://xxx.lanl.gov/abs/1112.3112}{{\tt arXiv:1112.3112}}].

\bibitem{NuFit}
M.~Gonzalez-Garcia, M.~Maltoni, J.~Salvado, and T.~Schwetz, {\it {Global fit to
  three neutrino mixing: critical look at present precision}},  {\em JHEP} {\bf
  1212} (2012) 123, [\href{http://xxx.lanl.gov/abs/1209.3023}{{\tt
  arXiv:1209.3023}}].

\bibitem{BilenkiBetaBeta}
S.~Bilenky and C.~Giunti, {\it {Neutrinoless double-beta decay: A brief
  review}},  {\em Mod.Phys.Lett.} {\bf A27} (2012) 1230015,
  [\href{http://xxx.lanl.gov/abs/1203.5250}{{\tt arXiv:1203.5250}}].

\bibitem{CP}
N.~Cosme, J.~Frere, and L.~Lopez~Honorez, {\it {CP violation from dimensional
  reduction: Examples in (4+1)-dimensions}},  {\em Phys.Rev.} {\bf D68} (2003)
  096001, [\href{http://xxx.lanl.gov/abs/hep-ph/0207024}{{\tt
  hep-ph/0207024}}].

\bibitem{B_mesons}
M.~Libanov, N.~Nemkov, E.~Nugaev, and I.~Timiryasov, {\it {Heavy-meson physics
  and flavour violation with a single generation}},  {\em JHEP} {\bf 1208}
  (2012) 136, [\href{http://xxx.lanl.gov/abs/1207.0746}{{\tt
  arXiv:1207.0746}}].

\bibitem{Cheng:2002iz}
H.-C. Cheng, K.~T. Matchev, and M.~Schmaltz, {\it {Radiative corrections to
  Kaluza-Klein masses}},  {\em Phys.Rev.} {\bf D66} (2002) 036005,
  [\href{http://xxx.lanl.gov/abs/hep-ph/0204342}{{\tt hep-ph/0204342}}].

\bibitem{bounds}
M.~Baak et~al.,
{\it {Updated Status of the Global Electroweak Fit and Constraints on New
Physics}}, {\em Eur.Phys.J.}  {\bf C72} (2012) 2003,
  [\href{http://xxx.lanl.gov/abs/1107.0975}{{\tt arXiv:1107.0975}}].

\bibitem{BsGammaLoops}
U.~Haisch and A.~Weiler,
{\it {Bound on minimal universal extra dimensions from $\bar B \to
  X(s)\gamma$}}, {\em Phys.Rev.}  {\bf D76} (2007) 034014,
  [\href{http://xxx.lanl.gov/abs/hep-ph/0703064}{{\tt hep-ph/0703064}}].

\bibitem{Rodejohann}
W.~Rodejohann, {\it {Neutrino-less Double Beta Decay and Particle Physics}},
  {\em Int.J.Mod.Phys.} {\bf E20} (2011) 1833--1930,
  [\href{http://xxx.lanl.gov/abs/1106.1334}{{\tt arXiv:1106.1334}}].

\bibitem{Gerda}
{\bf GERDA} Collaboration, K.~Ackermann et~al., {\it {The GERDA experiment for
  the search of $0\nu\beta\beta$ decay in $^{76}$Ge}},  {\em Eur.Phys.J.} {\bf
  C73} (2013) 2330, [\href{http://xxx.lanl.gov/abs/1212.4067}{{\tt
  arXiv:1212.4067}}].

\end{thebibliography}

\providecommand{\href}[2]{#2}\begingroup\raggedright

\end{document}